Proceedings of « **Haptic, Audio, Visual Environments and Games** » (HAVE 2009), pp 54-57. Politecnico di Milano november 7-8, 2009 (Lecco, Italy). Supported by the IEEE Instrumentation and Measurement Society (this document is the author file)

# Coaching the Wii

*Evaluation of a physical training experiment assisted by a video game*

Hanneton S., Varenne A.
Laboratoire de Neurophysique et Physiologie
UMR 8119 CNRS and Paris Descartes University
Paris, France
sylvain.hanneton@parisdescartes.fr

Hanneton, S.
UFR STAPS – Movement Sciences Department
Paris Descartes University
Paris, France

*Abstract*— **Aging or sedentary behavior can decrease motor capabilities causing a loss of autonomy. Prevention or re-adaptation programs that involve practice of physical activities can be precious tools to fight against this phenomenon. "Serious" video game have the potential to help people to train their body mainly due to the immersion of the participant in a motivating interaction with virtual environments. We propose here to discuss the results of a preliminary study that evaluated a training program using the well-known WiiFit game and Wii balance board device in participants of different ages. Our results showed that participants were satisfied with the program and that they progressed in their level of performance. The most important observation of this study, however was that the presence of a real human coach is necessary in particular for senior participants, for security reasons but also to help them to deal with difficulties with immersive situations.**

*serious game; physical activity; training; coaching; virtual environment, immersion (key words)*

## I. Introduction

Aging and/or sedentary behavior are known to cause metabolic pathologies (diabetes, cardio-vascular pathologies , obesity etc...) but also physical deterioration of sensory-motor behavior. Disorders of motion, balance and coordination, loss of reactivity and low energy expenditure are the expression of this physical deterioration. Physical activity is known to be efficient in the prevention or limitation these problems. If the amount of practice of 'static' video games is known to be correlated with sedentary behaviour [1], new interaction devices, like the Wiimote or Wii balance board, equipped with motion or force sensors have allowed the development of a kind of "serious" gaming that involves motion of the entire body and energy expenditure. Non static video games developed for solitary or family practice of physical activities at home have emerged recently and may substitute real activity. Even if the commercial target of the game industry is a family game for all ages [2], health and aging professionals consider that the practice of such games may be a valuable alternative or complement to classical care methods for their patients [3]. However, apart from medical case studies of injuries caused by extensive practice of the games (see [4] for instance), scientific publications regarding (1) the study of conditions in which the game has to be practiced, (2) the physical benefits of this kind of training have been little studied [5][6][7][8][9]. Consequently, we propose to present and discuss the results of an experimental study that involved four participants of different ages returning to physical activities via ten sessions of practice of the Wii Fit game.

The aims of this study are :

- to experiment a short program of personal but supervised physical training for participants of different ages

- to provide an estimate of the changes in the level of performance of participants for eight different activities proposed by the WiiFit game.

- to establish if aging has an influence on the level of performance of the game and an influence on the way the participant interacts with the game and the coach.

- to question the ability for all participants to be able to take part of the immersive aspects of the game

- to discuss if this practice can be efficient and secure without the help of the coach.

## II. Methods

### A. Participants

Participants were four female volunteers who were aged respectively 23, 29, 36 and 60 years old (designated as P23, P29, P36 and P60). None of the participants had known physical or neurological pathologies. None of them carried out physical activity or played video games. Importantly, they had no previous experience of the Wii balance board. They gave informed consent for the experiment and filled in a short questionnaire at the end of the last session. Each participant was individually instructed by both the real coach and the virtual coach proposed by the Wii. Each participant had her own Mii, a character that is the avatar of the participant in the game. The game required the use of the Nintendo[TM] Wii game console, the wii balance board device. The Wiimote, the remote controller that is the common interface for the Wii console, was sometimes required and used either by the participant or the coach depending on the activity. The Wii balance board is a rectangular plate-form 42 centimeter in length and 25 centimeter of width. Four force sensors recorded the weight of the player and the trajectory of the center of pressure of the participant during the game. Two Wii were kindly lent by the Nintendo France Corporation.

## B. Training sessions and role of the coach

An intensive program of ten sessions with two or three sessions per week was carried out. Each session lasted between 30 to 45 minutes. The functions of the real coach were to (1) give complementary explanation concerning the rules of the games if required by the participant, (2) to ensure the security of the participant and particularly to prevent the risk of fall, (3) to prevent participants from using incorrect postures or movements or to cheating movements.

## C. Choice of activities

A variety of activities involving the whole body are available with the WiiFit game. Certain activities are accessible at the beginning of the game while others are unveiled when the player overstep certain performance levels. We chose to focus on a restricted panel of eight activities that we considered as accessible for every age (see table one). Activities relating to the control of balance are voluntary over represented for two reasons : (1) the balance board device is obviously particularly adapted for the practice of balance games and (2) one targeted application of such game is the rehabilitation of balance for instance for elderly people. We distinguish between activities that simply rely on the mimicking of a movement or posture and "funny" activities in which participant has to deal with even simple rules. All these activities were present at the beginning of the game. However, we allowed the participant to choose the last activity of the session in order to increase motivation. Each session ended with a particular activity : a physical test where the mean position of the center of gravity and the BMI are recorded by the Wii.

| posture or movement | | fun activities | |
|---|---|---|---|
| **Yoga** | **Fitness** | **Balance** | **Aerobic** |
| Abdominal respiration | Torsions (vertical and horizontal) | Headers (soccer) | Step class |
| Posture of the warrior (left or right) | | Skiing | |
| | | Ski jump | |
| | | Marbles | |

Table 1: Activities chosen for the ten training sessions.

## D. Estimation of the level of performance and final questionnaire

The outcome measures used in this study were based on the performance data provided by the Wii fit game: for each session we recorded the BMI, the mean position of the center of pressure (CP), the "WiiFit age", and the scores of the different activities the participant practiced. We used a Pearson correlation coefficient test between the dependent variables and session numbers to evaluate the existence of a significant change in scores across sessions..The aim of the final questionnaire was to try to measure the subjective impression and motivation of participants and the possible subjective benefit of the sessions. The questions were :

1. did you find that activities proposed by this game were difficult ? (yes/no answer)
2. What did the game bring to you ? (free answer)
3. What did you particularly notice during the sessions ? (free answer)
4. Did you have particular expectations about this training that were not satisfied ? (free answer)
5. Now that the sessions are finished, what are you considering to do next ? (free answer)

The coach was also instructed to note every interesting observation or remarks of participants. He also noted also the supplementary activity chosen by each participant at the end of each session.

## III. RESULTS

### A. Feasibility of the program

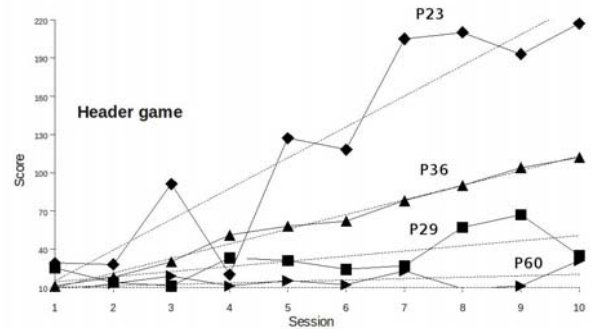

Figure 1: Changes in the score for the header game across sessions for the four participants. Dashed lines represent the linear approximation of the changes in scores.

The program was correctly achieved without any problems. All the participants were able to practice the chosen activities. A short pause was frequently asked by participants in the middle of the session. The first session lasted longer than expected because of the explanation require by the participant concerning the rules of the activities and the use of the interactive device.

### B. Changes in scores across sessions

There were no significant changes for the BMI or the WiiFit age across sessions for all the participants. The changes in scores across sessions was not always significant (see figure 1 for instance). Concerning the posture or movement activities, the scores of activities that change significantly were the abdominal respiration for two women (P29 and P36) and the posture of the warrior (right) for one women (P36). There were more significant changes for fun activities (table 2).

| Dependent variable or activity | | Participants | | | |
|---|---|---|---|---|---|
| | | P23 | P29 | P36 | P60 |
| Posture or movement activities | Body Mass Index (BMI) | | | | |
| | WiiFit Age | | | | |
| | Abdominal respiration | | X | X | |
| | Posture of the warrior (left) | | | | |
| | Posture of the warrior (right) | | | X | |
| | Vertical torsions | | | | |
| | Horizontal torsions | | | | |
| Fun activities | Headers (soccer) | X | X | X | |
| | Skiing | X | | | |
| | Ski jump | | | | |
| | Marbles | X | X | X | |
| | Step class | X | X | X | X |

*Table 2: List of dependant variables and chosen activities : crosses indicate a significant correlation ($\alpha=0.05$) between score and session number.*

The level of performance is not strictly dependent on age of the participant. However the oldest woman exhibited a level of performance systematically lower than the other participants.

*C. Results of the questionnaire*

Participants found that activities were not difficult (100%). Three participants reported that the activities gave them a better knowledge of their body. They said they did not have any particular expectation regarding their participation in the experiment except one participant who expected a greater level of cardio-training than she felt was provided by the game. Half of the participants felt that the sessions were too short. At the end of the program, participants said that they wished to start a physical activity (75 %) or/and to buy the game console (50 %).

*D. Virtual environment and usability*

The coach noticed real difficulties notably for the oldest participant to deal with the different virtual environments proposed by the game. Players are confronted with three kind of situations :

- a demonstration environment : the virtual coach demonstrates the appropriate movement that the player has to execute (yoga activities for instance)
- a game environment : the visual stimulation represents objects controlled by players without the intention of producing an immersive effect (marbles).
- An immersive environment where the players are supposed to represent themselves in the virtual environment with the possible help of an avatar (headers or skiing for instance).

Interestingly, the visual environment that caused the greatest difficulties was the demonstration environment : the oldest woman for instance hesitated to reproduce movements made by the virtual coach as in a mirror or inversely, and tried to rotate herself in order to reproduce the attitude of the coach but in doing so she turned her back to the screen. The help and explanations of the real coach were essential to allow this participant to adapt to the situation. One negative aspect of the immersive environment is that participants may have difficulties in both playing the game and ensure their own security in the real world. The plate-form of the Wii balance board is rather narrow and a fall is possible. This point was critical for the oldest participant : she had no difficulties in immersing herself in the game but was not able to do that and at the same time pay attention not to fall off the board.

IV. CONCLUSION

As mentioned before, this exploratory study was designed to establish if a program of physical training assisted by video games can be proposed to participants of different ages who are neither familiar with physical activities nor with the practice of video games. One other question that we wanted to address was the need for a real human coach during sessions of practice of the game. Even if the population of participant was small, this study gave precious indications concerning these questions.

First, the program was globally successful since all the participants found that it was pleasant and efficient, notably concerning their knowledge of their bodies. The changes in scores show a global progression among sessions, especially for the activities that present a fun component. However, it is difficult to make a statement to which extent the game logic and the Wii device are supporting the training since significant changes in the level of performance could be related only to the learnability of the game user interface and not to the improvement of sensory-motor skills. We did not observe any change of the BMI or of the WiiFit age but we suppose that this is due to the fact that the duration of the experiment was short (about five weeks) with activities that do not require a large energy expenditure. The program was motivating since participants expressed the wish to continue the physical stimulation..

However, age appears as an important factor that influence both the evolution of performance and the conditions in which the program has to be organized. The oldest participant did not show a significant improvement in scores except for the step class activity. But this significant improvement can be explained by a very bad level of performance for the first session due to difficulties in understanding the instructions and code used by the game to drive the steps of the player. Consequently, a similar program designed for seniors should last longer to provide a significant change in the level of performance. The contribution of the coach appears to be crucial for this population for both security and efficiency reasons. The difficulties of elderly people to maintain their attention both to interactions with the game and to their real environment and posture can be explained by cognitive impairments caused by aging. For instance, difficulties to in carrying out dual tasks are known to be common for elderly people [10]. Perhaps the old lady had difficulties to mentally follow the game and physically balance at the same time. The specific question of the behavior of seniors confronted with virtual environments is a point that has to be clearly addressed by the video game development community.

This work was a very informative preliminary study for a more ambitious longitudinal study concerning the use of the WiiFit game and of the Wii balance board. For a study involving people with known pathologies, more participants have to be recruited for a longer program (i..e. at least 10-12 weeks). The choice of activities and evaluation methods has to take into account gender differences in perception and spatial navigation abilities. Quantitative evaluation via some cognitive and sensory-motor tests has to be introduced before the first session and after the last one in order to establish if the changes in the scores displayed by the game are correlated with the improvement of cognitive and executive functions of the players [11].

This experience was also very informative concerning the eventual concrete use of the WiiFit game for rehabilitation programs in institutions for elderly people : we strongly recommend the presence of one or several coaches for both security and usability reasons.

ACKNOWLEDGMENT

We thank the Nintendo France Corporation for the lending of consoles.

REFERENCES

[1] M. Ballard, M. Gray, J. Reilly and M. Noggle. Correlates of video game screen amon males : body mass, physical activity and other media use. Eating Behaviors (2009, in press).

[2] E. T. Khoo, T. Merrit and A.D. Cheok. Designing physical and social intergenerational family entertainment. Interacting with computers 21:76-87 (2009).

[3] M. Aimonetti. Intérêt de la Wii pour les personnes âgées : oui à la Wii ! Neurologie Psychiatrie Geriatrie 9:63-64, Avril 2009.

[4] K.M. Hirpara, O.A. Abouazza. The "Wii Knee": A case of patellar dislocation secondary to computer video games. Injury Extra 39(3):86-87, March 2008.

[5] L. Lanningham-Foster, R. C. Foster, S. K. McCrady, Teresa B. Jensen, N. Mitre, James A. Levine. Activity-Promoting Video Games and Increased Energy Expenditure. The Journal of Pediatrics 154(6):819-823, June 2009.

[6] W. R. Boot, A. F. Kramer, D. J. Simons, M. Fabiani and G. Gratton. The effect of video game playing on attention, memory, and executive control. Acta Psychologica 129:387-398 (2008).

[7] T. Baranowski, R. Buday, D.I. Thompson and J. Baranowski. Paying for real : video games and stories for health-related behavior change. AM J Prev Med 34(1): 74-82 (2008).

[8] A.L. Betker, T. Szturm, Z. K. Moussavi, C. Nett. Video game-based exercises for balance rehabilitation : a single subject design. Arch Phys Med Rehabil 87:1141-1149 (2006).

[9] J.E. Deutsch, M. Borbely, J. Filler, K. Huhn, P. Guarrera-Bowlby. Use of a low-cost, commercially available gaming console (Wii) for rehabilitation of an adolescent with cerebral palsy. Physical Therapy 88(10):1196-207.(2008)

[10] P. Verhaeghen and J. Cerella. Aging, executive control, and attention: a review of meta-analyses. Neuroscience & Biobehavioral Reviews 26(7):849-857 (2002).

[11] C. Basak, W. R. Boot, M. W. Voss, A. F. Kramer. Can training in a real-time strategy video game attenuate cognitive decline in older adults? Psychology and Aging 23(4):765-777 (2008).